# Transparent Conducting Electrodes based on 1D and 2D Ag Nanogratings for Organic Photovoltaics

**Beibei Zeng, Zakya H. Kafafi[*], Filbert J. Bartoli[**]**
*Department of Electrical and Computer Engineering; Center for Photonics and Nanoelectronics, Lehigh University, Bethlehem, PA 18015*
[*]*zhk209@lehigh.edu*; [**]*fjb205@lehigh.edu*

**Abstract** The optical and electrical properties of optically-thin one-dimensional (1D) Ag nanogratings and two-dimensional (2D) Ag nanogrids are studied, and their use as transparent electrodes in organic photovoltaics are explored. A large *broadband* and *polarization-insensitive* optical absorption enhancement in the organic light-harvesting layers is theoretically and numerically demonstrated using either single-layer 2D Ag nanogrids or two perpendicular 1D Ag nanogratings, and is attributed to the excitation of surface plasmon resonances and plasmonic cavity modes. Total photon absorption enhancements of *150%* and *200%* are achieved for the optimized single-layer 2D Ag nanogrids and double (top and bottom) perpendicular 1D Ag nanogratings, respectively.

**Keywords**: organic photovoltaics, nanogratings, transparent electrodes

## 1. Introduction

Various light trapping strategies have recently been explored to enhance the optical absorption without increasing the thickness of the light-harvesting layers in photovoltaics (OPVs) [1-5]. One promising approach incorporates plasmonic nanostructures, for which the decay length of plasmonic modes at the metal/organic semiconductor interface is of the same order of magnitude as the thickness of the organic light-harvesting layer(s) [6-12]. Several metallic nanostructure geometries, including randomly distributed nanoparticles and periodically patterned nanoaperture arrays, have been proposed to increase the *optical but not the physical* thickness of the active light-harvesting layer(s) in OPVs [1, 3, 6].

Indium tin oxide (ITO) is currently the most widely used transparent conducting electrode (TCE) for OPVs, due to its high electrical conductivity and optical transparency [13]. However, the limited availability and increasing cost of ITO, its incompatibility with flexible substrates, and poor mechanical and chemical stability reduce its attractiveness for use in OPVs [14-16]. Hence, new materials and approaches have been suggested as alternatives to ITO electrodes, including macroscopic metallic grids, nanowires, randomly perforated metal films, carbon nanotubes, and graphene [17-21]. One-dimensional (1D) metallic nanogratings and two-dimensional (2D) metallic nanogrids have been theoretically and experimentally investigated and shown to possess high optical transmission and electrical conductivity, making them particularly attractive as TCEs [22-25].

OPVs with ultrathin 1D Ag nanogratings TCEs have the potential for broadband absorption enhancement and have recently been shown to achieve stronger optical absorption and higher power conversion efficiencies (PCEs) than those with ITO electrodes [10-12]. A *total absorption enhancement* of *50%* in the organic light-harvesting layers was reported for OPVs with 1D Ag nanogratings TCEs [11], and *67%* for OPVs with two parallel 1D Ag nanogratings used as the top and bottom electrodes [10]. Enhancement of the short circuit current density J*sc* (*40%*) and PCE (*35%*) was achieved when 1D Ag nanogratings were used as electrodes in molecular OPVs [12]. However, only one specific polarization of the incident light was able to excite the plasmonic modes and achieve strong light-trapping effects [23, 24]. The simultaneous optimization of the absorption enhancement for the incident light with different polarizations is difficult for 1D metallic nanostructures. Consequently, it is critical to design plasmonic TCEs that will lead to *broadband* and *polarization-insensitive* optical absorption enhancement in the organic light-harvesting layers. Recently, we reported that plasmonic TCEs consisting of 2D metallic nanogrids provide polarization-independent light-trapping effects and lead to further enhancement of the optical absorption in the organic light-harvesting layers [23]. However, SPRs are not excited (and the electromagnetic fields are not enhanced) in the central part of the 2D Ag nanogrids for both polarizations, since the electric field is parallel to one of the grating directions. Hence, the central areas of the 2D Ag nanogrids do not contribute to the optical absorption enhancement in the adjacent active layers. To further enhance the optical absorption in the adjacent active layers, it is necessary to design new plasmonic nanostructures in order to excite SPRs over the whole area, including the central part. In the current work, sandwiching the OPV structure between one top and one bottom 1D metallic nanogratings with perpendicular directions is proposed to excite plasmonic modes for both polarizations, potentially leading to larger polarization-independent absorption enhancement. In the following, we report a systematic study of the optical and electrical properties of ultrathin 1D Ag nanogratings and 2D Ag nanogrids, explore their use as plasmonic TCEs in OPVs, and optimize absorption enhancement as a function of nanograting and nanogrid's thickness, linewidth, and period.

## 2. Electrical and optical properties of ultrathin 1D Ag nanogratings and 2D Ag nanogrids

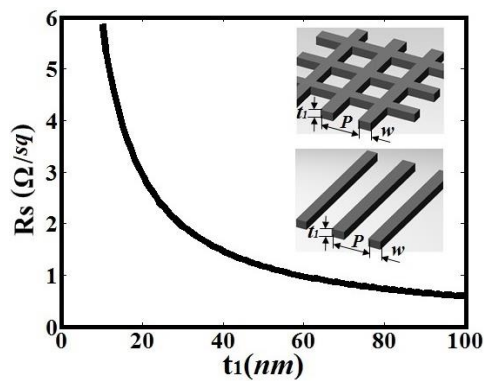

Fig. 1. Electrical sheet resistance of 1D Ag nanogratings or 2D Ag nanogrids (insets show the geometry with period *P=300nm*, line-width *w=70nm*) electrodes as a function of film thickness $t_1$.

The proposed plasmonic Ag nanogratings or nanogrids TCEs are much thinner than typical metal films used in classical optical studies. As the thickness of the metal film approaches its skin depth in the visible electromagnetic spectral region (tens of nanometers) [26, 27], it exhibits electronic and optical properties quite different from those of a typical optically-thick metal films [22]. The electrical properties of an electrode is usually described by its sheet resistance $R_s$. For 1D Ag nanogratings or 2D Ag nanogrids with linewidth *w*, thickness $t_1$, and period $P$ (see the insets of Fig. 1), $R_s$ can be calculated using the expression $R_s = (\rho/t_1)(P/w)$ [22], where $\rho = 1.58 \times 10^{-8} \Omega \cdot m$ is the electrical resistivity of a bulk Ag film [28]. $R_s$ changes from *0.67* to *6.7$\Omega/sq$* as the thickness $t_1$ of the 1D Ag nanogratings or 2D Ag nanogrids decreases from *100* to *10nm* (*P=300nm* and *w=70nm*, these values are the optimum for the largest absorption enhancement in OPV devices, as will be discussed in the following sections), as depicted in Fig. 1. These theoretically estimated values for $R_s$ agree reasonably well with previous experimental measurements and are well below those of ITO thin film (*10~20$\Omega/sq$*) used as transparent electrodes [24].

In addition to excellent electrical conductivity, high optical transmission through the ultrathin Ag nanogratings or nanogrids is another important requirement for TCE applications. Three-dimensional (3D) finite-difference time-domain (FDTD) simulations were used to investigate the underlying physical mechanisms that determine the electronic and optical properties of ultrathin Ag nanogratings and nanogrids [29]. These simulations assume that the 2D Ag nanogrids with linewidth $w$, thickness $t_1$ and period $P$ are directly placed on a glass substrate. Fig. 2(a) shows 2D maps of the calculated transmission spectra for the 2D Ag nanogrids as a function of the period and incident wavelength when the linewidth and thickness of the 2D nanogrids are fixed at *w=70nm* and *$t_1$=30nm*, respectively. The decreased transmission in region 1 (*550nm<λ<900nm* and *100nm<P<160nm*) of Fig. 2(a) is caused by the high optical reflection, due to the cutoff of the propagating electromagnetic waves through the vertically-oriented Ag-air-Ag waveguide [22].

The low transmission in region 2 (*350nm<λ<500nm* and *100nm<P<400nm*) is attributed mainly to the intrinsic absorption in the 2D Ag nanogrids, as shown in Fig. 2(b). Three different mechanisms may be responsible for the enhanced absorption in the 2D Ag nanogrids. For the ultrathin 2D Ag nanogrids, two single-interface surface plasmon polariton (SI-SPP) modes at the top and bottom of the Ag/dielectric interfaces would interact with each other and lead to coupled long-range and short-range SPP (LR- and SR-SPP) modes. The dispersion relations for LR- and SR-SPP modes in optically-thin continuous metal film can be described by the following equation [30, 31]:

$$\tanh(k_2 t_1)(\varepsilon_{d1}\varepsilon_{d2}k_2^2 + \varepsilon_m^2 k_1 k_3) + \varepsilon_m k_2 (\varepsilon_{d1} k_3 + \varepsilon_{d2} k_1) = 0 \qquad (1)$$

Here $k_1^2 = k_{spp}^2 - \varepsilon_{d1} k_0^2$, $k_2^2 = k_{spp}^2 - \varepsilon_m k_0^2$, $k_3^2 = k_{spp}^2 - \varepsilon_{d2} k_0^2$, $k_0 = \omega/c$ and $t_1$ is the thickness of the metal film. $\varepsilon_{d1}$, $\varepsilon_{d2}$ and $\varepsilon_m$ are the dielectric constants of air, glass and Ag, respectively. For the *30nm*-thick Ag film with an asymmetric geometry ($\varepsilon_{d1} < \varepsilon_{d2}$), only the strongly damped SRSPP modes exist with anti-symmetric $E_z$ field patterns at the top (Air/Ag) and bottom (glass/Ag) interfaces [26]. It is also known that the momentum mismatch between the SPP modes and free space light can be bridged by the reciprocal vectors of the periodic nanostructures $k_G = mG_x + nG_y$, where $|G_x| = |G_y| = 2\pi/P$, $m$ and $n$ are integers and $\theta$ is the incident angle as shown in this expression [31]:

$$k_{spp} = k_0 \sin\theta + mG_x + nG_y \qquad (2)$$

The dispersion relation of the SR-SPP modes can be obtained by substituting Eq. (2) into Eq. (1). The black solid curve in Fig. 2(b) refers to the analytical dispersion of the SRSPP modes excited by the lowest order reciprocal vectors {(*m,n*)=(*1,0*),(*-1,0*),(*0,1*),(*0,-1*)}, which agrees well with the absorption spectra where the period is in the range of *100nm<P<220nm*, since the dispersion relation of ultrathin Ag nanogrids with large duty cycles (linewith/period) can approximate that of continuous metal films.

As the period *P* continues to increase from *220nm* to *300nm* while keeping the line-width of the Ag nanogrids fixed at *w=70nm*, the distance between the two adjacent Ag strips becomes larger, resulting in weaker coupling between them [24, 26]. Therefore, the absorption in the 2D Ag nanogrids with the period varying from *220nm to 300nm* is attributed to the excitation of the localized-SPRs (LSPRs), supported by the individual metallic nanostructures [3]. The spectral positions of the absorption band remain almost unchanged as the period increases from *220nm* to *300nm*. This is due to the fact that the resonance wavelength of the LSPRs is primarily determined by the geometry of the individual Ag strips, which are fixed at *w=70nm* and *$t_1$=30nm*, respectively. When the momentum matching condition is satisfied ($\omega/c \cdot \sqrt{\varepsilon_d} = k_0 \sin\theta + mG_x + nG_y$, $\varepsilon_d$ represents the dielectric constant of air or glass), the typical Rayleigh-Wood Anomaly (RA) occurs at the Ag/glass and air/Ag interfaces [33], respectively, as shown by the black dash-dotted and dashed lines in Fig. 2(b). The ±1 diffraction orders of the incident light are scattered parallel to the metal surfaces, giving rise to the strong optical absorption in the Ag nanogrids [24].

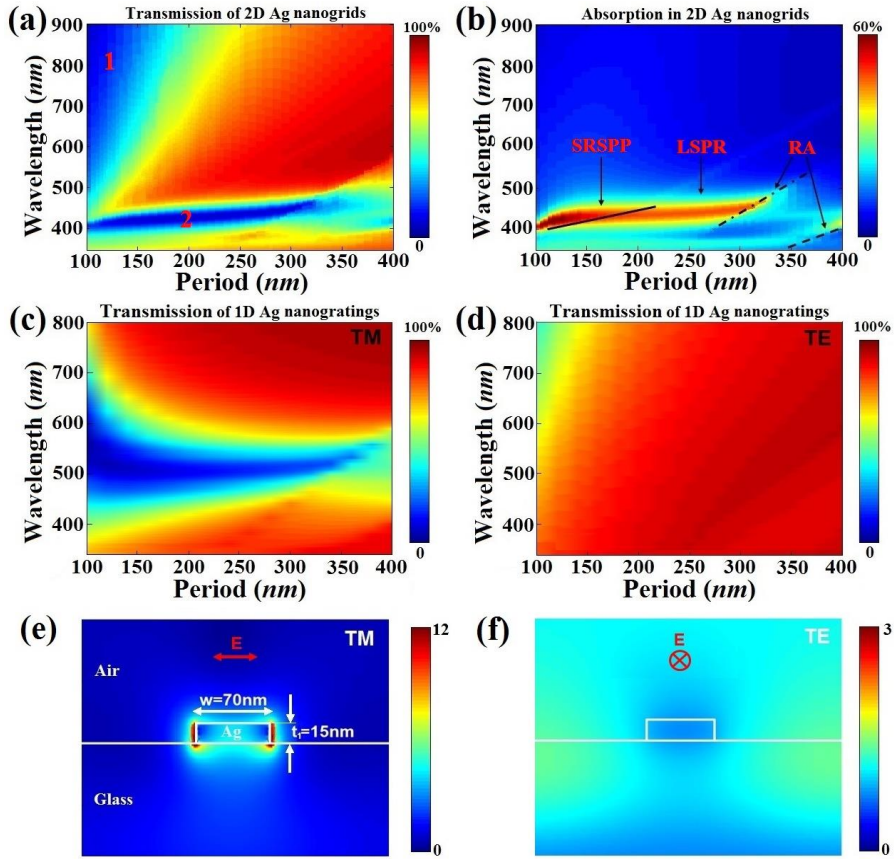

Fig. 2. Color maps of the calculated optical transmission (a) and absorption (b) spectra for 2D Ag nanogrids as a function of the period *P* and incident wavelength when line-width *w* and thickness *$t_1$* are fixed at *70nm* and *30nm*, respectively. The black solid and dash-dotted (dashed) curves in (b) refer to the analytical dispersions of SRSPP modes and Rayleigh-Wood anomaly at the Ag/glass (Ag/air) interface, respectively. Color maps of the calculated optical transmission of 1D Ag nanogratings (thickness *$t_1$=15nm* and line-width *w=70nm*) as a function of the period and incident wavelength, under (c) TM and (d) TE polarizations, respectively. The electric field distribution in the *15nm*-thick 1D Ag nanogratings (linewidth *w=70nm*, period *P=300nm*) under (e) TM and (f) TE polarizations, respectively, when the incident wavelength is *500nm*.

3D FDTD simulations were also used to investigate the optical properties of ultrathin 1D Ag nanogratings under different polarization conditions. Fig. 2(c) and (d) show color maps of the calculated optical transmission spectra for the 1D Ag nanogratings (thickness $t_1=15nm$ and linewidth $w=70nm$, these parameters are the optimum for absorption enhancement in OPV devices, as will be discussed in the following sections) as a function of its period and the incident wavelength, under TM and TE polarizations, respectively. In Fig. 2(c), the low transmission in the region of *400nm<λ<700nm* and *100nm<P<150nm* is caused by the high optical reflection, which is due to the cutoff of the propagating electromagnetic waves through 1D Ag nanogratings with small air openings (*30~80nm*). For 1D Ag nanogratings with *P>150nm*, the transmission minimum is located around *λ=500nm*, and is attributed mainly to the intrinsic absorption of the Ag nanogratings. The small decay length (tens of nanometers) of the electromagnetic field depicted in Fig. 2(e) demonstrates that the resonance mode is due to LSPRs (electromagnetic field is highly localized around the nanostructures) in the *15nm*-thick 1D Ag nanogratings (linewidth *w=70nm* and period *P=300nm*) at *500nm*. The TE-polarized incident light (Fig. 2(f)) transmits through the ultrathin 1D Ag nanogratings (thickness $t_1=15nm$ and line-width *w=70nm*) with high transmission (>90%) over the entire visible region. Fig. 2(f) clearly shows that there is no resonant electromagnetic mode under TE polarization, in which case the incident light simply passes through the air openings between Ag strips.

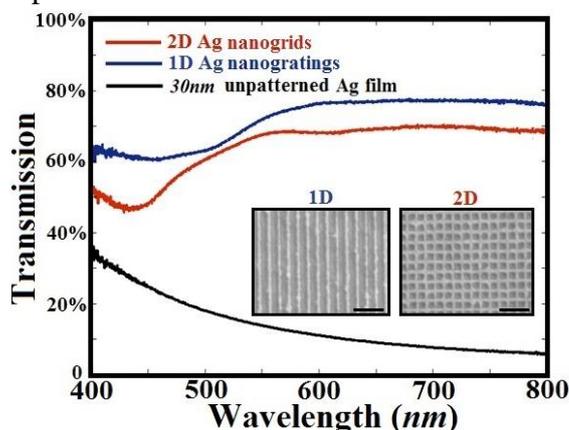

Fig. 3. Measured optical transmission through 1D Ag nanogratings and 2D Ag nanogrids (blue and red curves) with *P=300nm*, *w=70nm*, and $t_1=30nm$, and *30nm*-thick unpatterned flat Ag film (black curve). Insets show SEM images of the fabricated 1D Ag nanogratings and 2D Ag nanogrids. Scale bar, *1μm*.

Focused ion beam milling (FEI Dual-Beam system 235) was used to fabricate 1D Ag nanogratings and 2D Ag nanogrids with period *P=300nm* and linewidth *w=70nm* on a *30nm*-thick Ag film, deposited by E-beam evaporation (Indel system) onto a glass slide (Fisherbrand). An Olympus X81 inverted microscope system was employed to measure the optical transmission of the fabricated metallic nanostructures. An unpolarized white light beam from a *100W* halogen lamp illuminated the sample surface through a condenser. The field and aperture diaphragms of the microscope were both closed to obtain a nearly collimated light beam. The transmitted light was collected through a *40×* objective lens (numerical aperture, NA=*0.6*), and then coupled to a fiber-based portable spectrometer (Ocean Optics, USB4000). *50* spectrum frames were averaged with an integration time of $\tau=5msec$ for each spectrum, and then normalized to the reference spectrum of an open aperture with the same area as nanogratings or nanogrids. Fig. 3 shows the normalized transmission spectra for the 2D Ag nanogrids (red curve), 1D Ag nanogratings (blue curve), and a *30nm*-thick unpatterned flat Ag film (black curve). Although there are differences between the experimental and the numerical results due to fabrication and measurement uncertainties [23], there are obvious transmission valleys observed at the wavelength of *450nm* for both the 2D Ag nanogrids and 1D Ag nanogratings, which are attributed to the excitation of LSPRs in the individual Ag strips [24]. The optical transmission of the 2D Ag nanogrids and 1D Ag nanogratings is much higher than that of the unpatterned flat Ag film over the entire visible region (*400~800nm*). While the optical transmission of the 1D Ag nanogratings is slightly higher than that of the 2D Ag nanogrids, the transmission near the resonance wavelength of *450nm* is lower for the 2D Ag nanogrids, due to a higher Ag strip density of 2D Ag nanogrids (in two directions) that can contribute to LSPRs under different polarizations. The higher mode density of LSPRs can generate stronger field intensity and thus greater absorption enhancement in the adjacent OPV active layers.

## 3. Optical absorption enhancement in OPVs with 1D Ag nanogratings and 2D Ag nanogrids electrodes

Fig. 4(a) and (b) illustrate schematic diagrams of molecular OPVs with an ultrathin 2D Ag nanogrids electrode and two perpendicular top and bottom 1D Ag nanogratings electrodes, respectively. In Fig. 4(a), the *30nm*-thick 2D Ag nanogrids electrode (linewidth *w* and period *P*) is positioned on a glass substrate and embedded in the poly(3,4-ethylenedioxythiophene):poly(styrenesulfonate) (PEDOT:PSS) film. Next are the light-harvesting layers, consisting of a *10nm*-thick electron donor copper phthalocyanine (CuPc) layer and a *10nm*-thick electron acceptor perylene tetracarboxylic bisbenzimidazole (PTCBI) layer, followed by an *8nm*-thick bathocuproine (BCP) layer that further transports electrons to the *80nm*-thick bottom Ag electrode. In Fig. 4(b), the *15nm*-thick top 1D Ag nanogratings electrode (line-width $w_1$ and period $P_1$) is positioned on a glass substrate and embedded in the PEDOT:PSS film, and is covered by the organic light-harvesting layers (*10nm*-thick CuPc and *10nm*-thick PTCBI layers). This is followed by a *60nm*-thick bottom 1D Ag nanograting (linewidth $w_2$ and period $P_2$) where the grating direction is perpendicular to the top 1D Ag nanograting embedded in the BCP layer, and on top of the *80nm*-thick bottom Ag electrode.

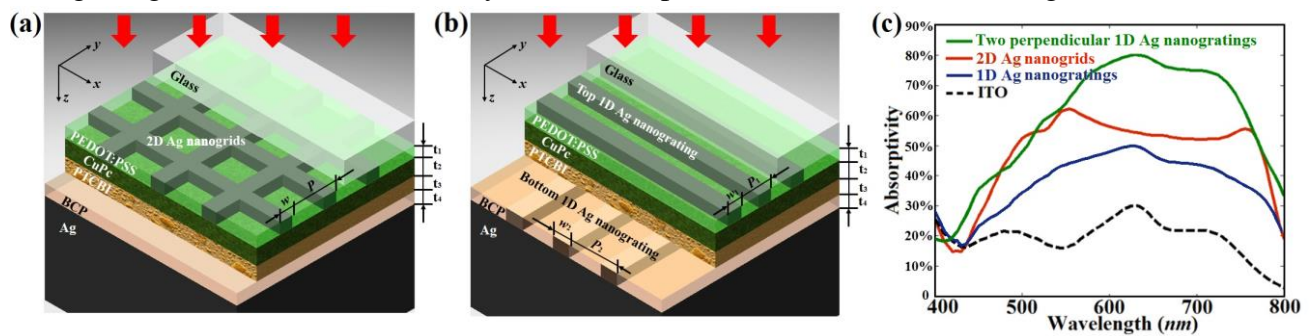

Fig. 4. Schematic diagrams of the proposed molecular OPVs with (a) a top 2D Ag nanogrids electrode, and (b) two perpendicular 1D Ag nanogratings electrodes. (c) Calculated optical absorption spectrum *A(λ)* in the organic active light-harvesting layers (CuPc:PTCBI) with a 100*nm*-thick ITO electrode (black dashed curve), 1D Ag nanogratings ($t_1=30nm$, $w=70nm$, and $P=300nm$, blue curve), 2D Ag nanogrids ($t_1=30nm$, $w=70nm$, and $P=300nm$, red curve), and two perpendicular 1D Ag nanogratings ($w_1=w_2=70nm$, $P_1=P_2=300nm$, green curve).

The 3D FDTD numerical method used to calculate the optical absorption in the CuPc:PTCBI light-harvesting layers allows the direct incorporation of the experimental optical constants of PEDOT:PSS, CuPc, PTCBI, BCP and Ag [34,35]. The optical absorption spectrum $A(\lambda)$ in the active layers is determined via FDTD simulations by calculating the difference between the optical power incident on and transmitted through the active light-harvesting layers of the OPV device, and then normalizing it to the incident optical power [8,9]. The simulations distinguish between the light that is absorbed or back-scattered by the plasmonic nanostructures and the forward propagating light. Only the light propagating past the plasmonic nanogratings and into the active light-harvesting layer is employed in calculations of the absorption of the organic layer. In the 3D FDTD simulation, a unit cell (consisting of one Ag strip with the active layers and Ag back reflectors) was used with periodic boundary conditions in the *x* and *y* directions ($x=y=P$) to simulate an infinite array of periodic nanogratings or nanogrids. Perfectly matched layer boundary conditions were used in the vertical *z* direction ($z=\pm5\mu m$) to prevent unphysical scattering at the edge of the simulation box. A coarse mesh size of 4*nm* was used over the whole simulation box. In the area of great interest (e.g. Ag nanostructures, thin active layers), the mesh size was decreased to 2*nm*, which is small enough to ensure the simulation accuracy. Fig. 4(c) shows the calculated optical absorption spectra $A(\lambda)$ of the OPV active layers with two perpendicular 1D Ag nanogratings ($w_1=w_2=70nm$, $P_1=P_2=300nm$, green curve), *30nm*-thick 2D Ag nanogrids ($w=70nm$, $P=300nm$, red curve) and 1D Ag nanogratings ($w=70nm$, $P=300nm$, blue curve), and *100nm*-thick ITO electrodes (black dashed curve) under TM and TE polarized incident light, respectively. These geometric parameters are the optimum for achieving the strongest absorption enhancement in the organic active light-harvesting layers, as will be discussed in the following sections. The optical absorption $A(\lambda)$ in the active layers with the two perpendicular 1D Ag nanogratings is the strongest throughout most of the visible spectral region. The optical absorption $A(\lambda)$ with the 2D Ag nanogrids is much stronger and broader than that with the 1D Ag nanogratings or the ITO electrode. The broadband absorption enhancement with the 2D Ag nanogrids and 1D Ag nanogratings can be attributed to the excitation of SPRs, and the formation of plasmonic cavity modes between the top 1D Ag nanogratings or 2D Ag nanogrids and the bottom Ag back reflector electrode, which can significantly confine and enhance the electromagnetic field in the ultrathin organic light-harvesting layers [9-12].

In order to elucidate the underlying physical mechanisms, different cross-sections (x-z, x-y, and y-z) of the electric field distribution for the OPVs with the top 1D Ag nanogratings, 2D Ag nanogrids and the two perpendicular 1D Ag nanogratings electrodes are depicted in Fig. 5(a-c), respectively, under different polarizations (electric fields along x and y axis) at the resonance wavelength $\lambda=600nm$. Fig. 5(a,i), (b,i) and (b,iii) show that strong electric field enhancement in the CuPc:PTCBI active layers occurs for the 1D Ag nanogratings (electric field along x axis) or 2D Ag nanogrids (electric field along x and y axis) at the resonance wavelength of *600nm*. Due to the polarization-dependent excitation of SPRs in 1D Ag nanogratings, the field intensity in the active layers could only be enhanced for the incident light with the electric field perpendicular to the grating direction, as shown in the Fig. 5(a,i) and (a,ii). In addition, Fig. 5(a, iii) and (a, iv) show that the field intensity in the active layers is slightly suppressed when the electric field of the incident light is parallel to the grating direction, since neither SPRs or waveguide modes are supported, and part of the incident light is blocked by the 1D Ag nanogratings [11]. On the other hand, the x-z and y-z cross-sections of the electric fields in Fig. 5(b,i) and (b,iii) are identical, due to the polarization-independent excitation of SPRs in the 2D Ag nanogrids. Thus, as shown in Fig. 4(c), the optical absorption $A(\lambda)$ in OPVs with the 1D Ag nanogratings is weaker than that with the 2D Ag nanogrids, in which the SPRs and plasmonic cavity modes can be excited under both x and y polarizations.

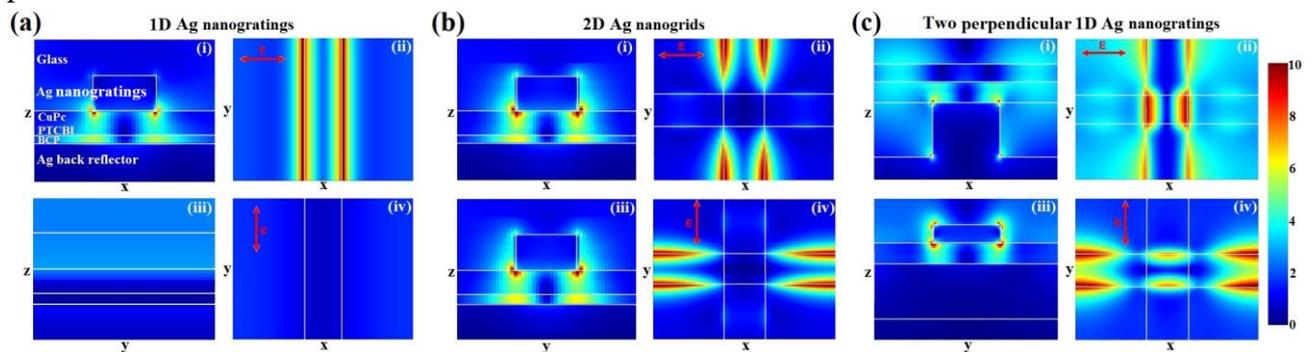

Fig. 5. Different cross-sections (x-z, x-y, and y-z) of the calculated electric field distributions for OPVs with (a) 1D Ag nanogratings, (b) 2D Ag nanogrids, and (c) two perpendicular 1D Ag nanogratings under different polarizations (electric fields along x or y axis) at the resonance wavelength $\lambda=600nm$.

The optical absorption in the OPV active layers with the 2D Ag nanogrids is stronger and broader than that with the 1D Ag nanogratings electrode. However, for each specific polarization (e.g., electric field along x or y axis), the x-y cross-sections in Fig. 5(b,ii) and (b,iv) show that the SPRs are not excited and thus the electric fields are not enhanced in the central part of the 2D Ag nanogrids under both polarizations, since the electric field is parallel to one of the grating directions. The central areas of the 2D Ag nanogrids do not contribute to the optical absorption enhancement in the adjacent active layers. To further enhance the optical absorption in the adjacent active layers, it is necessary to design a different geometry for the Ag nanostructures in order to excite SPRs over the whole area, including the central part. Fig. 5(c,i) and (c,iii) clearly demonstrate that the electric fields are confined and enhanced at the bottom and top 1D Ag nanogratings under x and y polarizations, respectively. The incident light with the electric fields along the x-axis, as shown in Fig. 5(c,i) and (c,ii), is highly transmitted through the top ultrathin 1D Ag nanogratings (grating direction along x-axis) (Fig. 2(d)), and then arrives at the bottom 1D Ag nanogratings (the grating direction is along y-axis, perpendicular with the top 1D Ag nanogratings), resulting in the excitation of the SPRs and absorption enhancement in the adjacent active layers. For the incident light with the electric fields along the y-axis (Fig. 5(c,iii) and (c,iv)), SPRs can be excited in the top 1D Ag nanogratings (along the x-axis). Due to the separately excited SPRs at the top and bottom 1D Ag nanogratings under different polarizations, the electric field is enhanced over the whole gratings area, including the central area under both polarizations, as depicted in Fig. 5(c,ii) and (c,iv). Therefore, the top and bottom perpendicular 1D Ag nanogratings electrodes provide much stronger optical absorption enhancement in the organic active light-harvesting layers than that calculated for the OPV active layers with the single layer 2D Ag nanogrids top electrode. Note that both LSPRs and the plasmonic cavity modes can be excited in OPVs with the top 1D Ag nanogratings, 2D Ag nanogrids and the two perpendicular 1D Ag nanogratings electrodes, as is clearly shown in Fig. 5 (a-c), in which these two modes couple with each other and contribute together to the absorption enhancement in the active light-harvesting layers.

## 4. Geometric optimizations of single-layer 2D Ag nanogrids and two perpendicular top and bottom 1D Ag nanogratings

*4.1 Geometric optimization for 2D Ag nanogrids*

In order to maximize the overall optical absorption in the OPV light-harvesting layers (CuPc:PTCBI) with the 2D Ag nanogrids electrode, the geometric parameters (period, linewidth, and thickness) of the 2D Ag nanogrids need to be optimized. The solar photon flux density is defined as $\Phi_s(\lambda)=S(\lambda)\cdot\lambda/hc$, where $S(\lambda)$ is the AM1.5 solar irradiance spectrum. The absorbed photon flux density, $A(\lambda)\cdot\Phi_s(\lambda)$, for the OPV device is determined via FDTD simulations, where $A(\lambda)$ is the optical absorption in the organic light-harvesting layers. The total photon absorption $A_{photon}$, which represents the fraction of the total solar photon flux density absorbed by the organic active layers, can be calculated using the equation [6]:

$$A_{photon} = \int_{\lambda_{min}}^{\lambda_{max}} A(\lambda)\cdot\Phi_s(\lambda)d\lambda \bigg/ \int_{\lambda_{min}}^{\lambda_{max}} \Phi_s(\lambda)d\lambda \qquad (3)$$

The visible spectral region of interest is *400~800nm* for the CuPc:PTCBI active layers [10-12]. $A_{photon-ref}$ refers to the total photon absorption in the reference OPV structure using a *100nm*-thick ITO electrode. Fig. 6(a) gives the total photon absorption $A_{photon}$ and its enhancement $(A_{photon}/A_{photon-ref}-1)\cdot 100\%$ as a function of the period (*P*) and thickness ($t_1$) of the 2D Ag nanogrids while the linewidth is fixed at *w=70nm*. $A_{photon}$ with thinner 2D Ag nanogrids is larger as the period changes from *100nm* to *150nm*, since thinner Ag nanogrids have higher optical transmission. $A_{photon}$ continues to increase as the period of the *30nm*-thick Ag nanogrids increases up to *300nm*, reaching a maximum value of *0.49*5. Fig. 6(b) shows $A_{photon}$ and its enhancement as a function of the period *P* and line-width *w* when the thickness of the Ag nanogrids is fixed $t_1=30nm$. The maximum $A_{photon}$ occurs when the linewidth is equal to *70nm* and the period is *300nm*. The maximum $A_{photon}$ of *0.495* for OPVs with the optimized 2D Ag nanogrids represents an enhancement of *150%* compared to that of the reference OPV structure using an ITO electrode. In OPV devices with the optimized plasmonic nanostructures, the enhanced electromagnetic field in the active layers is aligned with their absorption profiles. Note that $A_{photon}$ is greatly enhanced for a large range of values for the period *P* and/or thickness $t_1$ of the 2D Ag nanogrids, which provides good tolerance for future fabrication of these OPV nanostructures.

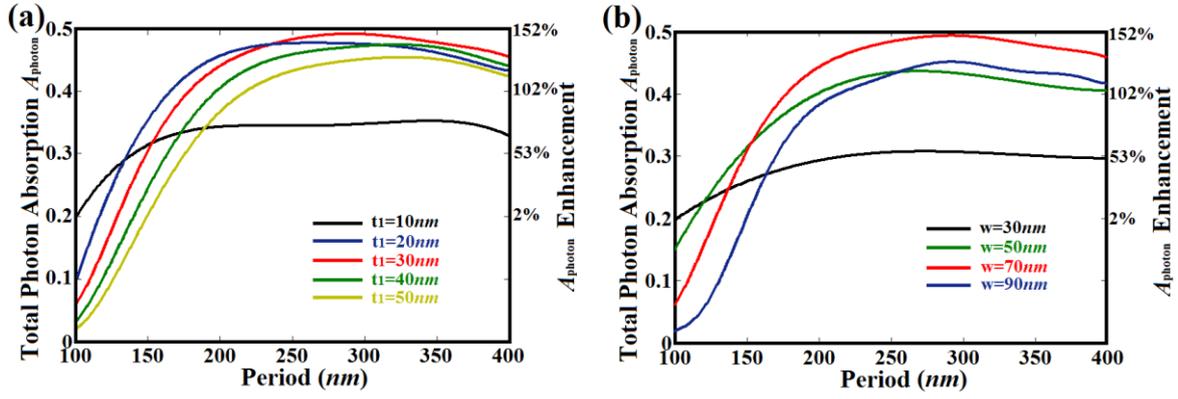

Fig. 6. Total photon absorption $A_{photon}$ and its enhancement $(A_{photon}/A_{photon-ref}-1)\cdot 100\%$ in the organic active layers of OPVs as a function of (a) period $P$ and thickness $t_1$ of the 2D Ag nanogrids, when the line-width $w$ is fixed at *70nm*; (b) period $P$ and line-width $w$, when the thickness $t_1$ is *30nm* (all other geometric parameters are the same as that in Fig. 3(a)).

It is worth noting that the total photon absorption $A_{photon}$ and its enhancement with the optimized 2D Ag nanogrids, *0.495* and *150%*, respectively, are even greater than that (*0.45* and *128%*) with a much more complex double 2D plasmonic nanostructure [9], consisting of top Ag nanodisc arrays and bottom Ag nanohole arrays sandwiching the OPV organic layers. It is reasonable to expect that the OPV device with the top 2D Ag nanogrids electrode would be much easier to fabricate that the more complex 2D nanostructure composed of top Ag nanodisc arrays and bottom Ag nanohole arrays. If the AM1.5 solar irradiance spectrum $S(\lambda)$ is used as a weighting factor instead of the photon flux density $\Phi_s(\lambda)$, the overall optical absorption will be defined as the *total absorptivity* $A'_{photon}=\int_{\lambda_{min}}^{\lambda_{max}} A(\lambda)\cdot S(\lambda)d\lambda / \int_{\lambda_{min}}^{\lambda_{max}} S(\lambda)d\lambda$ [10,11], where $A(\lambda)$ is the optical absorption in the organic light-harvesting layers. And the *total absorption enhancement* is $(A'_{photon}/A'_{photon-ref}-1)\cdot 100\%$. The *total absorptivity* and *total absorption enhancement* in the organic active layers with the proposed 2D Ag nanogrids are *0.48* and *141%*, respectively, which are much larger than *0.24* and *50%* predicted for that with the 1D Ag nanogratings [11]. It is also larger than the *total absorption enhancement* of *67%* for OPVs with parallel top and bottom 1D Ag nanogratings electrodes [10].

### *4.2 Geometric optimization of two perpendicular top and bottom 1D Ag nanogratings*

The geometric parameters of the two perpendicular top and bottom 1D Ag nanogratings need also to be optimized in order to maximize the overall optical absorption enhancement in the active light-harvesting layers. Due to the complex scheme shown in Fig. 4(b), the geometric parameters of the bottom 1D Ag nanogratings are first examined. Fig. 7(a) and (b) show a plot of the total photon absorption $A_{photon}$ and its enhancement $(A_{photon}/A_{photon-ref}-1)\cdot 100\%$ in an OPV with a bottom 1D Ag nanogratings and a top ITO electrode, as a function of thickness $t_4$ and line-width $w_2$ of the bottom 1D Ag nanogratings (period $P_2$ is fixed at *300nm*); or period $P_2$ (the thickness and linewidth are fixed at $t_4$ =*60nm* and $w_2$=*70nm*, respectively). In Fig. 7(a), $A_{photon}$ is increased as the thickness $t_4$ changes from *30nm* to *60nm*, and the linewidth $w_2$ varies from *20nm* to *70nm*. $A_{photon}$ reaches a maximum when the thickness and linewidth are kept at $t_4$=*60nm* and $w_2$=*70nm*, respectively. Fig. 7(b) shows that the maximum $A_{photon}$ occurs when the period $P_2$ is equal to *300nm*.

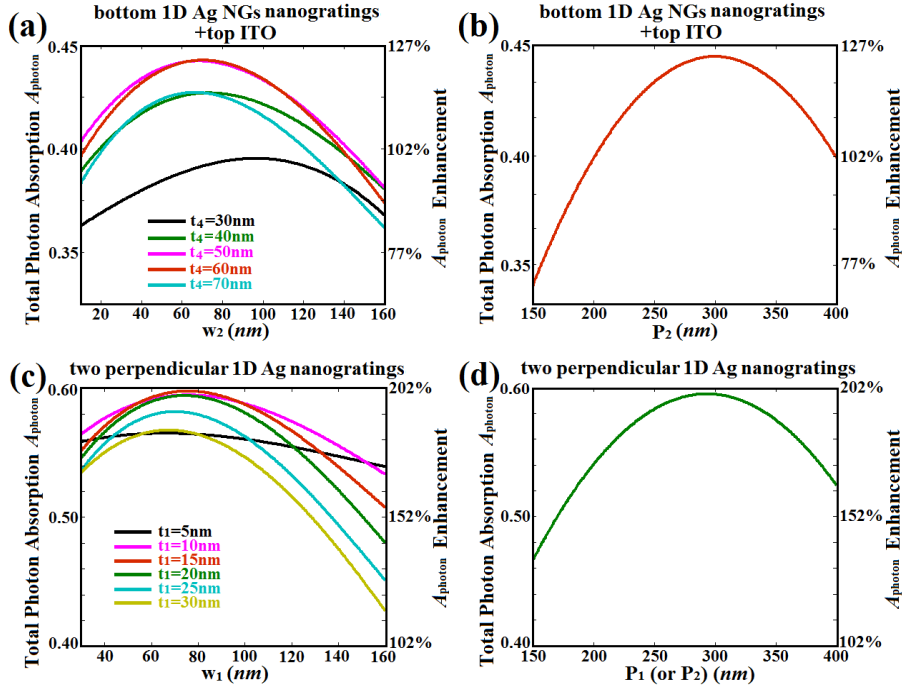

Fig. 7. Total photon absorption $A_{photon}$ and its enhancement in OPVs with bottom 1D Ag nanogratings and top ITO electrodes, as a function of (a) thickness $t_4$ and line-width $w_2$ of bottom 1D Ag nanogratings, when the period $P_2$ is fixed at *300nm*, (b) period $P_2$, when the thickness and line-width are fixed at $t_4 =60nm$ and $w_2=70nm$, respectively; and in OPVs with two perpendicular top and bottom 1D Ag nanogratings (bottom 1D Ag nanogratings with the optimized parameters $t_4 =60nm$ and $w_2=70nm$), as a function of (c) thickness $t_1$ and line-width $w_1$ of the top 1D Ag nanogratings, when the period is fixed at *300nm*, (d) period $P_1$ (or $P_2$), when the thickness and linewidth are $t_1 =15nm$ $w_1=70nm$, respectively.

The total photon absorption $A_{photon}$ and its enhancement are calculated in OPVs with two perpendicular top and bottom 1D Ag nanogratings (the bottom 1D Ag nanogratings with the optimized parameters $t_4 =60nm$ and $w_2=70nm$), as a function of thickness $t_1$ and linewidth $w_1$ of the top 1D Ag nanogratings (period is fixed at *300nm*); or period $P_1$ (or $P_2$) (the thickness and linewidth are $t_1=15nm$ $w_1=70nm$, respectively), as shown in Fig. 7(c) and (d). Fig. 7(c) shows that $A_{photon}$ is increased as the thickness $t_1$ increases from *5nm* to *15nm*, and the linewidth $w_1$ varies from *30nm* to *70nm*, respectively. $A_{photon}$ decreases as the thickness $t_1$ or the linewidth $w_1$ increases further, since the thicker or wider top Ag nanostrips result in lower optical transmission. $A_{photon}$ reaches a maximum value of *0.595* when the thickness and the linewidth of the top 1D Ag nanogratings are $t_1=15nm$ and $w_1=70nm$, respectively, corresponding to an $A_{photon}$ enhancement of *200%*. Fig. 7(d) shows that $A_{photon}$ and its enhancement as a function of the period $P_1$ (or $P_2$) of top (or bottom) 1D Ag nanogratings when the thickness and linewidth of top and bottom 1D Ag nanogratings are set at their optimized values. The maximum $A_{photon}$ occurs when the period is set at *300nm*. $A_{photon}$ is greatly enhanced over a large range of geometric parameters of the top and bottom 1D Ag nanogratings, indicating once more a great tolerance for future fabrication of these nanostructures. Note that $A_{photon}$ and its enhancement with the optimized two perpendicular 1D Ag nanogratings, *0.595* and *200%*, respectively, are much greater than those achieved in the previously suggested OPV design with the 2D Ag nanogrids (*0.495* and *150%*) [23]. However, the fabrication of the OPV devices with the two perpendicular top and bottom 1D

Ag nanogratings would probably be more challenging than that with a single layer top 2D Ag nanogrids [36].

## 5. Angular dependence of the optical absorption enhancement

In order to prevent the usage of expensive mechanical tracking systems that are often used to align the solar panel surface normal to the incident light, the solar cell needs a broad angular response [37-40]. Thus, it is necessary to investigate the angular dependence of the optical absorption enhancement in the proposed OPV nanostructures. The total photon absorption $A_{photon}$ and its enhancement $(A_{photon}/A_{photon-ref}-1)\cdot 100\%$ for OPVs with the optimized single-layer 2D Ag nanogrids decreased from *0.495* and *150%* for normal incidence ($\theta=0^0$) to *0.152* and *30.6%* for an incident angles of $\theta=60^0$, as red solid and blue dashed curves as shown in Fig. 8(a), respectively. Compared with our previous designs (polymeric OPVs with a top Ag nanodisc array and a flat Ag back reflector electrode) [6], the $A_{photon}$ enhancement with the 2D Ag nanogrids is stronger either under normal incidence (e.g. *150% v.s. 31.2%* at $\theta=0^0$) or under larger incident angles (e.g. *30.6% v.s. 28.7%* at $\theta=60^0$). In addition, the total photon absorption $A_{photon}$ and its enhancement for OPVs with the optimized two perpendicular top and bottom 1D Ag nanogratings decreased from *0.595* and *200%* for normal incidence to *0.164* and *41%* under *60º* incident angles (red solid and blue dashed curves in Fig. 8(b)). Compared with OPVs with the 2D Ag nanogrids, the $A_{photon}$ enhancement with the perpendicular top and bottom 1D Ag nanogratings is further increased either under normal incidence (e.g. *200% v.s. 150%* at $\theta=0^0$) or large incident angles (e.g. *41% v.s. 30.6%* at $\theta=60^0$).

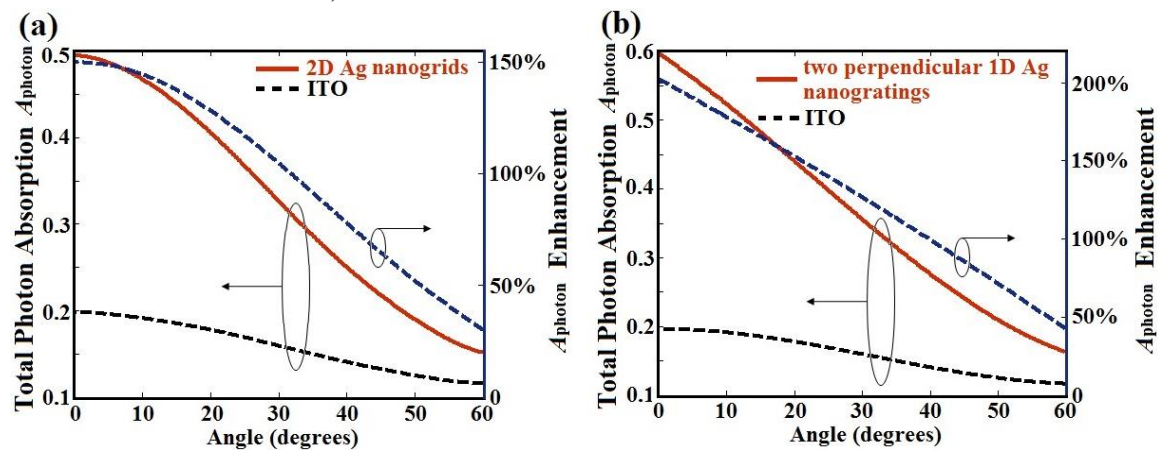

Fig. 8. Angular dependence of the total photon absorption for molecular OPVs with optimized (a) single-layer 2D Ag nanogrids; and (b) two perpendicular 1D Ag nanogratings electrodes ($A_{photon}$, red solid curves), and ITO electrodes ($A_{photon-ref}$, black dashed curves), and the corresponding $A_{photon}$ enhancement (blue dashed curves).

## 6. Conclusion

In summary, we have systematically studied the optical and electrical properties of novel plasmonic transparent electrode consisting of ultrathin 1D Ag nanogratings and 2D Ag nanogrids, with a calculated sheet resistance <10 Ω/sq. The underlying physical mechanisms that determine the optical properties of the ultrathin 1D Ag nanogratings and 2D Ag nanogrids have been investigated and delineated. Strong optical absorption enhancement in the OPV organic light-harvesting layers with 2D nanogrids, and two perpendicular top and bottom 1D Ag nanogratings electrodes have been calculated. The total photon absorption $A_{photon}$ was increased to quite high values of *0.495* and *0.595* for the optimized single-layer 2D Ag nanogrids and two-layer perpendicular top and bottom 1D Ag nanogratings, respectively. This represents enhancements in optical absorption of *150%* and *200%*, respectively, compared to that of a reference OPV with a conventional ITO electrode. The fabrication of the OPV device with the two perpendicular top and bottom 1D Ag nanogratings, which provides the strongest absorption enhancement, may prove to be more challenging than that with a single layer 2D Ag nanogrids. These design principles are quite general and can be extended to other organic, inorganic, and organic/inorganic hybrid optoelectronic devices with thin active layers that are adjacent to the plasmonic nanostructures. Since plasmonic resonances are very sensitive to the geometric parameters of the metallic nanostructures and the dielectric constants of the surrounding materials [41-46], careful consideration is required for each specific design and material used.

## Author contributions

B. Z. conceived the idea, and is responsible for the theoretical design, simulation, and experimental demonstration. Z. H. K. and F. J. B. directed the project, and contributed to the interpretation of the results and the writing of the manuscript.